# GaN-based Bipolar Cascade Lasers with 25nm wide Quantum Wells: Simulation and Analysis

J. Piprek,[a] G. Muziol,[b] M. Siekacz,[b] and C. Skierbiszewski [b]
[a] NUSOD Institute LLC, Newark, DE 19714-7204, United States, E-mail: piprek@nusod.org
[b] Institute of High Pressure Physics, Polish Academy of Sciences, Sokołowska 29/37, 01-142 Warsaw, Poland

*Abstract*— Utilizing self-consistent numerical simulation, we analyze internal device physics, performance limitations, and optimization options for a unique laser design with multiple active regions separated by tunnel junctions, featuring surprisingly wide quantum wells. Contrary to common assumptions, these quantum wells are revealed to allow for perfect screening of the strong built-in polarization field, while optical gain is provided by higher quantum levels. However, internal absorption, low p-cladding conductivity, and self-heating are shown to strongly limit the laser performance.

## I. INTRODUCTION

Intense worldwide efforts have been focusing on efficiency improvements of GaN-based light emitters.[1] Among the most intriguing proposals is the cascading of multiple active regions with tunnel junctions in between. Such multi-junction devices were previously demonstrated for several types of light emitters, including GaAs-based lasers [2] and GaSb-based light-emitting diodes (LEDs).[3] Dual-wavelength GaN-based LEDs combined two different active regions based on the same concept.[4] In all these cases, electrons and holes are recycled by tunneling and used repeatedly for photon generation, allowing for quantum efficiencies above 100%.[5] Vertical stacks of laser diodes are especially attractive for pulse mode applications, such as gas sensing, printing and environment pollution control, or light detection and ranging (LIDAR) in cartography, automotive and industrial systems.[6] They are a cheap and viable alternative to laser bar arrays. Junction cascades provide much simpler coupling of the light output with external optics than from lateral arrays of lasers, since the spatial separation between vertical devices is two orders of magnitude smaller. The simultaneous operation of a cascade of *n* lasers increases the slope efficiency of the full device *n*-times, which makes high-power lasing conditions accessible for smaller currents. In addition, the level of catastrophic optical damage is *n*-times higher in comparison to a single laser.

However, the AlGaInN material system makes it difficult to fabricate multi-junction devices with traditional metal organic vapor-phase epitaxy (MOVPE) so that the expected high slope efficiencies have not been reported yet. The primary reason lies in the passivation of Mg acceptors by hydrogen and the need for post-growth activation of p-type conductivity through thermal annealing, which removes hydrogen. On the other hand, plasma-assisted molecular beam epitaxy (PAMBE), which is a hydrogen-free technology, recently led to GaN-based bipolar cascade lasers with more than 100% differential quantum efficiency,[7] which is the quantum efficiency above lasing threshold. A rather unique feature of this laser is the employment of very thick InGaN quantum wells (QWs, see Tab. 1) which enables the screening of interface polarization charges by low-level QW carriers.[8] In good agreement with measurements, we here analyze the internal physics of this novel laser by self-consistent numerical simulation and evaluate performance limitations as well as design optimization options.

TABLE 1: EPITAXIAL LASER STRUCTURE

| Layer | Material | Thickness | Doping / $10^{18}$cm$^{-3}$ |
|---|---|---|---|
| TJ | $In_{0.02}Ga_{0.98}N$ | 20 nm | Si:40 |
| TJ | $In_{0.17}Ga_{0.83}N$ | 5 nm | Si:180 |
| TJ | $In_{0.17}Ga_{0.83}N$ | 5 nm | Mg:100 |
| TJ | $In_{0.02}Ga_{0.98}N$ | 60 nm | Mg:50 |
| cladding | GaN | 500 nm | Mg:1 |
| EBL | $Al_{0.14}Ga_{0.86}N$ | 20 nm | Mg:20 |
| waveguide | $In_{0.08}Ga_{0.92}N$ | 60 nm | - |
| top QW | $In_{0.18}Ga_{0.82}N$ | 25 nm | - |
| waveguide | $In_{0.08}Ga_{0.92}N$ | 80 nm | - |
| cladding | GaN | 300 nm | Si:1 |
| cladding | GaN | 200 nm | Si:40 |
| TJ | $In_{0.02}Ga_{0.98}N$ | 20 nm | Si:40 |
| TJ | $In_{0.17}Ga_{0.83}N$ | 5 nm | Si:180 |
| TJ | $In_{0.17}Ga_{0.83}N$ | 5 nm | Mg:100 |
| TJ | $In_{0.02}Ga_{0.98}N$ | 60 nm | Mg:50 |
| cladding | $Al_{0.05}Ga_{0.95}N$ | 400 nm | Mg:1 |
| waveguide | GaN | 100 nm | Mg:1 |
| EBL | $Al_{0.14}Ga_{0.86}N$ | 20 nm | Mg:20 |
| waveguide | $In_{0.04}Ga_{0.96}N$ | 110 nm | - |
| bottom QW | $In_{0.17}Ga_{0.83}N$ | 25 nm | - |
| waveguide | $In_{0.04}Ga_{0.96}N$ | 110 nm | - |
| waveguide | GaN:Si | 100 nm | Si:1 |
| cladding | $Al_{0.05}Ga_{0.95}N$ | 700 nm | Si:1 |

TJ – tunnel junction, EBL – electron blocker layer, QW – quantum well

## II. DEVICE DETAILS

The reference device is a blue light emitting laser diode featuring two InGaN single-quantum well active regions that are separated by an InGaN tunnel junction.[7] A second tunnel junction is grown on top for uniform carrier injection. The full layer structure of the active region is given in Tab. 1. The cleaved-facet cavity is 1mm long and the 15μm wide ridge is etched through both QWs. A SiO$_2$ passivation layer is deposited outside of the ridge and a standard Ti/Al/Ni/Au metallization is evaporated on both sides of the laser structure. The vertical waveguide structure and the two lasing modes are illustrated in Fig. 1. For characterization purposes, the two lasers were designed slightly different which results in two independent lasing modes.

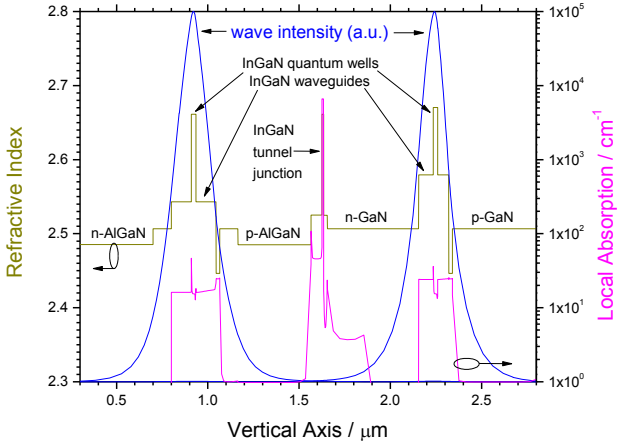

Fig. 1. Vertical profiles of refractive index, local absorption, and laser modes. The top tunnel junction is excluded for clarity, also in the following figures.

### III. MODELS AND MATERIAL PARAMETERS

Our customized laser model is based on the PICS3D simulation software.[9] It self-consistently computes carrier transport, the wurtzite electron band structure of strained InGaN quantum wells, stimulated photon emission, multi-mode wave guiding, and self-heating. The transport model includes drift and diffusion of electrons and holes, built-in polarization, Fermi statistics, thermionic emission at hetero-interfaces, interband tunneling, as well as all relevant radiative and non-radiative recombination mechanisms. Schrödinger and Poisson equations are solved iteratively in order to account for QW deformations (quantum-confined Stark effect). The large Mg ionization energy in p-doped layers is considered as well as Poole-Frenkel ionization enhancement by the local electric field. The thermal model calculates the internal heat power distribution generated by current flow, carrier recombination, and photon absorption. More details on the employed laser models can be found elsewhere.[10,11,12]

The threshold current for lasing operation is mainly controlled by the QW Auger recombination coefficient C for which widely scattered values are reported.[1] Within the reported range we find that $C = 6.1 \times 10^{-30}$ cm$^6$/s delivers good agreement with the measurement (see below). Defect-related QW Shockley-Read-Hall (SRH) recombination has negligible impact assuming a SRH lifetime of 20 ns. However, the laser performance is also limited by internal absorption. We here adopt a first-principle model [13] resulting in an absorption cross section of about $0.6 \times 10^{-18}$ cm$^2$ for free electrons. Hole-related absorption including non-ionized acceptors exhibits a cross section of about $0.9 \times 10^{-18}$ cm$^2$ for GaN. Alloy scattering raises the hole-related absorption cross-section to about $1.2 \times 10^{-18}$ cm$^2$. The highly doped tunnel junctions exhibit strong inter-band absorption of 6600/cm but contribute only a small fraction to the modal loss due to their location near minimum wave intensity (Fig. 1). However, the undoped InGaN waveguide layers cause significant modal loss, as identified in a previous study of similar lasers.[14] This is mainly attributed to electron transitions between the band tails formed by alloy disorder. As the waveguide band gap approaches the lasing wavelength with rising indium content, the band tails deepen and cause stronger absorption. We here employ the measured absorption of 16/cm for In$_{0.04}$Ga$_{0.96}$N and 24/cm for In$_{0.08}$Ga$_{0.92}$N.[14] The total absorption profile is shown in Fig. 1 at 2A current injection, including free carrier loss.

### IV. RESULTS AND ANALYSIS

The simulated energy band diagram is shown in Fig. 2 for an injection current of 2A. Electrons are injected from the left-hand side into the conduction band, recombine with holes inside the bottom QW, and continue inside the valence band towards the tunnel junction. After tunneling, electrons travel again inside the conduction band to the top QW where they can generate a second photon by recombining with holes arriving from the right-hand side (for clarity, the top tunnel junction is not shown here). In each laser section, electrons loose about 6eV of their energy, mainly inside the QW by generating a photon or by non-radiative recombination. Another significant energy loss of more than 2eV occurs inside the lowly p-doped cladding layers. The tunnel junction cascade doubles the total energy loss to more than 12eV. Correspondingly, the calculated bias is 12.3V at 2A. At lasing threshold near 0.5A, the calculated bias is 8V. It is much lower than the measured threshold bias of 13.6V because the simulation assumes a perfect tunnel junction with rectangular doping profiles that are impossible to achieve in reality. A detailed tunnel junction analysis and optimization is published elsewhere.[15]

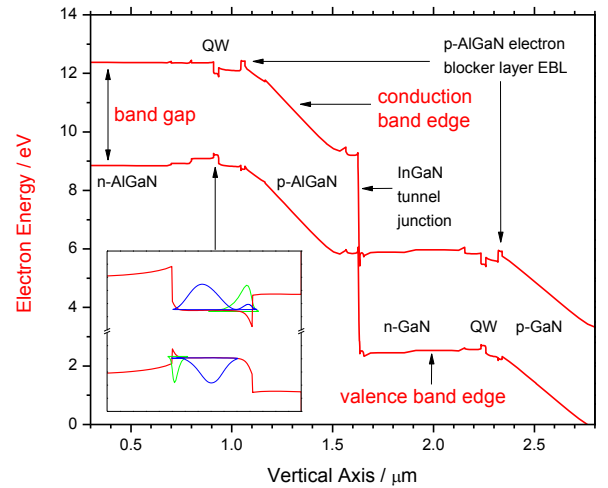

Fig. 2. Energy band diagram calculated at 2A current. The inset shows the first two QW energy levels and the corresponding wave functions.

The unusually thick QWs lead to a somewhat counter-intuitive phenomenon. GaN-based light emitters typically feature very thin QWs to minimize the separation of electrons and holes by the strong built-in QW polarization field. This separation is even stronger in our wide QWs as illustrated by the green wave-functions in the inset of Fig. 2. The lowest energy levels are confined at the QW edges, thereby eliminating the overlap of the corresponding wave-functions as well as any photon generation by radiative recombination between these two levels. However, the carriers located in these lowest quantum states completely screen the built-in polarization charges at the QW interfaces, leading to flat band edges in the center of the QW. Radiative transitions are enabled by higher quantum levels. The calculated wave-functions of the second QW levels are shown by the blue lines in Fig. 2. They exhibit a strong overlap and produce most of the optical gain in this device. Wider QWs allow for lasing at lower carrier densities and therefore reduce the current loss caused by Auger recombination and carrier leakage, which are both known to strongly rise with the QW carrier density.

Since the two lasers exhibit a slightly different design (cf. Tab. 1), their performance also differs. In good agreement with the measurement, the top laser turns on first (green line in Fig. 3). Its optical confinement factor is $\Gamma=9.7\%$ and the modal loss $\alpha_i=14.2$/cm. The bottom laser experiences lower modal loss (12.3/cm) but also less optical confinement (8.8%) so that its threshold current is somewhat higher (blue line in Fig. 3). In both cases, the absorption loss is dominated by the InGaN waveguide layers (cf. Fig. 1). The calculated total power (red line in Fig. 3) agrees quite well with the measurement (symbols). This simulation neglects self-heating since the experiment was performed in pulsed laser operation. However, even short current pulses cause some internal temperature rise, as indicated by the measured sub-linearity at high current. Considering equal emission from both laser facets, the differential quantum efficiencies are 57.4% and 54.5% for bottom and top laser, respectively, adding up to $\eta_d = 111.9\%$. In other words, injected electron-hole pairs generate more than one emitted photon, as intended.

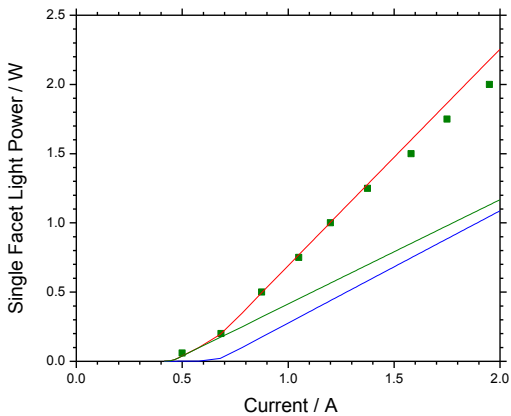

Fig. 3. Simulated light vs. current characteristics for bottom laser (blue) and top laser (green). The total power (red) agrees well with the measurement (symbols)

The differential quantum efficiency $\eta_d$ of laser diodes is known to be limited by carrier leakage from the QWs and/or by internal photon loss. The simulated current density profiles are plotted in Fig. 4 and reveal that minority carrier leakage currents are several orders of magnitude smaller than the current of majority carriers. The much discussed electron leakage is obviously very well suppressed by the p-AlGaN electron blocker layers in this laser. In other words, our differential quantum efficiency is only limited by internal absorption.

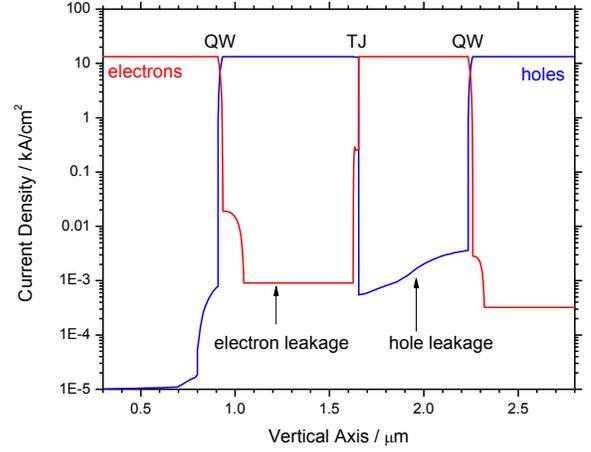

Fig. 4. Current density profiles for electrons and holes at 2A total current (QW – quantum well, TJ – tunnel junction).

The total power conversion efficiency (PCE) is the ratio of optical output power to electrical input power. It is hardly improved by the tunnel junction concept because a doubling of the output power is accompanied by a doubling of the electrical bias. In contrast to the record PCE values of more than 80% reported for GaN-based blue LEDs, the PCE of blue lasers is still below 50%.[16] This discrepancy is typically attributed to the low electrical conductivity of the p-doped cladding layers that are needed in laser diodes for wave guiding. We obtain a peak value of PCE=18% at 2A in our simulation, which is mainly limited by the very low doping of the p-side waveguide cladding layers (cf. Tab. 1 and Fig. 2). Due to the high Mg acceptor activation energy, the density of free holes is about two orders of magnitudes lower than the acceptor density. In the following, we therefore investigate cladding layer optimization options.

There are two main options for p-cladding layer improvements. First, higher Mg doping reduces the electrical resistance and the bias, but it increases photon absorption. Second, a lower p-cladding thickness also reduces the resistance but it moves the tunnel junction towards the lasing mode causing higher optical loss and lower power. Thus, both cases represent a trade-off between lower resistance and higher absorption, i.e., between bias reduction and power loss. For simplicity, we now simulate the combination of two identical lasers so that the results are easier to digest. The top laser in the original device is replaced by the design used for the bottom laser, including the p-AlGaN cladding layer. The simulated performance is close to the original. Table 2 lists the calculated results of thickness and doping variations.

TABLE 2:
RESULTS AT 2A CURRENT FOR P-CLADDING LAYERS OF
DIFFERENT THICKNESS AND DOPING DENSITY.

| Thickness nm | Doping $10^{18}$cm$^{-3}$ | Modal Loss cm$^{-1}$ | Lasing Power W | Bias V | PCE % |
|---|---|---|---|---|---|
| 400 | 1 | 12.3 | 2.31 | 12.81 | 18.1 |
| 350 | 1 | 12.5 | 2.31 | 12.19 | 19.0 |
| 300 | 1 | 12.8 | 2.26 | 11.59 | 19.5 |
| 250 | 1 | 13.3 | 2.22 | 11.02 | 20.1 |
| 200 | 1 | 14.4 | 2.12 | 10.45 | 20.3 |
| 150 | 1 | 16.6 | 1.95 | 9.52 | 20.5 |
| 100 | 1 | 20.7 | 1.68 | 9.08 | 18.5 |
| 400 | 2 | 12.4 | 2.30 | 11.14 | 20.7 |
| 400 | 5 | 12.8 | 2.27 | 9.72 | 23.4 |
| 400 | 10 | 13.3 | 2.22 | 9.03 | 24.6 |
| 400 | 15 | 13.9 | 2.17 | 8.74 | 24.8 |
| 400 | 20 | 14.6 | 2.12 | 8.57 | 24.7 |

PCE – power conversion efficiency

Due to the very low Mg doping, thickness reductions down to 150 nm result in slight PCE enhancements because the reduced bias has more impact than the increasing modal loss. The rising internal absorption is attributed to the rising tunnel junction overlap with the optical mode. Figure 5 illustrates the case with 150 nm p-cladding thickness. As expected from the low Mg doping, the cladding layer causes less than 0.1/cm of the modal loss while the tunnel junctions contribute more than 4/cm to the total loss of 16.6/cm per mode (red line in Fig. 5, the integrated loss represents an average loss per lasing mode as it is normalized by the total wave intensity of both modes). The tunnel junction contribution doubles with 100 nm thin p-cladding leading to a significant PCE decline (Tab. 2) so that the optimum p-AlGaN thickness appears to be near 150 nm. However, higher p-AlGaN conductivity would shift this optimum to larger values.

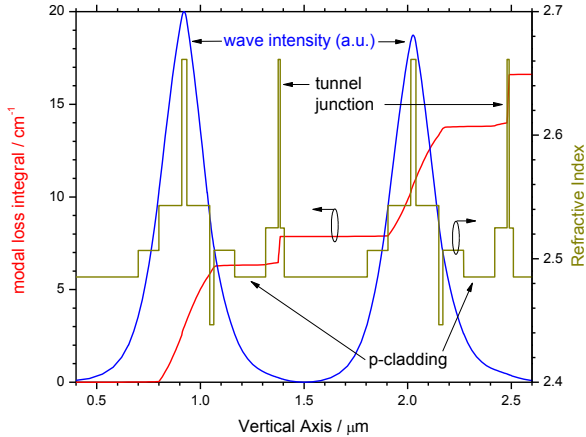

Fig. 5. Vertical profile of refractive index, wave intensity, and modal loss integral with 150nm p-AlGaN cladding. Note that a thin n-cladding layer was added on top to achieve identical waveguides.

Variation of the Mg doping in the p-cladding layer leads to a more significant efficiency enhancement up to PCE = 24.8% with $15 \times 10^{18}$cm$^{-3}$ Mg density (Tab. 2). The rising hole-related p-AlGaN absorption causes a relatively small increase of the total modal loss while the bias reduction is somewhat stronger than with cladding thickness reductions. In other words, doping enhancements promise higher efficiencies than thickness reductions.

Finally, we investigate obstacles to continuous-wave (CW) operation of these lasers, which has not been experimentally demonstrated yet. The thermal model is the same as in our previous study of GaN-based lasers which achieved excellent agreement with CW measurements.[12] We here select the design with the highest PCE from Tab. 2 which promises the lowest self-heating. The key parameter for CW laser simulations is the thermal resistance $R_{th}$ defined as the ratio of internal temperature rise to internal heat power. Values as low as $R_{th}$ = 6.6K/W have been reported for high-power GaN lasers.[17] Due to the relatively high thermal conductivity of III-nitrides, most of the total thermal resistance is caused by mounting and packaging. Therefore, we add an external thermal resistance to our laser simulation resulting in a total $R_{th}$ of 7K/W. The self-heating in quasi-CW operation with relatively long lasing pulses and high duty-cycles may be represented by a somewhat lower thermal resistance.[18] We therefore also show simulation results with $R_{th}$=5.4K/W and $R_{th}$=3.2K/W in Fig. 6 demonstrating the strong impact of this parameter. The peak output power rises from 1.3W at 7K/W to 7.8W at 3.2K/W. However, the much higher peak current still causes an internal temperature up to 500K which reduces the QW gain so that the QW carrier density needs to rise in order to maintain lasing (Fig. 7). This finding contradicts the common assumption of constant QW carrier density above threshold, which is only true with constant QW temperature.[12] The rising QW carrier density leads to rising carrier losses. Figure 8 compares the current consumption of all QW recombination processes for $R_{th}$=5.4K/W. The strongest increase is due to Auger recombination which consumes more current than stimulated emission at high temperature and eventually leads to the roll-off of the output power. Spontaneous photon emission and defect-related SRH recombination draw much less current. Current leakage from the QW also rises, but it remains negligibly small.

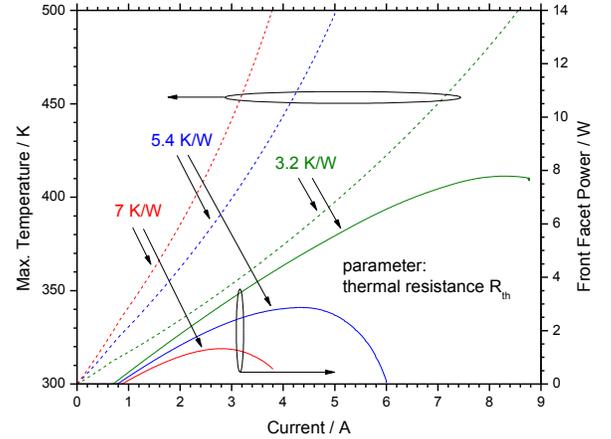

Fig. 6. Output power (solid lines) and maximum internal temperature (dashed lines) vs. current calculated for 3 different thermal resistances.

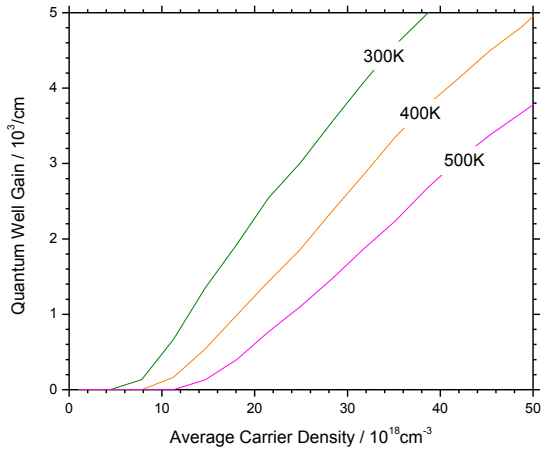

Fig. 7. Quantum well gain vs. average carrier density calculated for different temperatures.

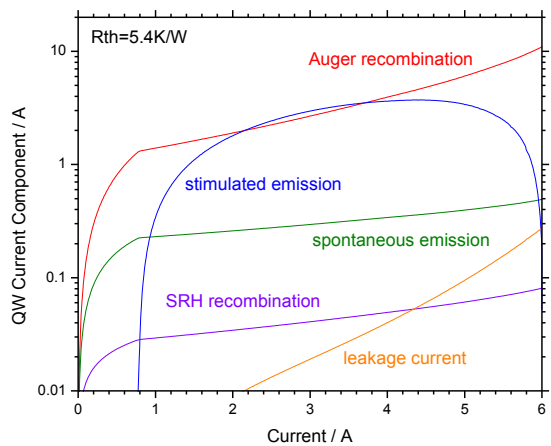

Fig. 8. Comparison of current components vs. total current. Due to carrier recycling by the tunnel junction, the sum of all components is twice the injection current.

Despite our wide quantum wells, Auger recombination remains a key obstacle towards CW operation of our laser design, due to strong self-heating. However, our simulations suggest that CW lasing can be achieved if two conditions are met. First, the minimization of the operating voltage, e.g., by optimized p-doping. Second, a proper mounting and packaging of the laser chip, which would ensure a total thermal resistance below 7K/W.

## V. SUMMARY

In summary, internal physical mechanisms and performance limitations of InGaN/GaN bipolar cascade lasers are revealed by self-consistent numerical simulation in good agreement with measurements. Contrary to common assumption, wide quantum wells are shown to allow for an effective screening of the internal polarization field by quasi two-dimensional carrier accumulation in the fundamental quantum levels. However, while exceeding 100%, it is found that the differential quantum efficiency is still severely limited by internal absorption. In addition, we show that the power conversion efficiency suffers from the low doping of the p-side cladding layers. Higher Mg acceptor densities promise a significant efficiency enhancement, despite sligthly stronger absorption. Furthermore, we investigate the thermal properties of our laser stack and demonstrate that CW operation requires a very small thermal resistance because strong self-heating leads to rising Auger recombination.

ACKNOWLEDGEMENTS

The contribution of G. M., M. S., and C. S was supported by funding from Narodowe Centrum Nauki (2019/35/D/ST3/03008), Narodowe Centrum Badan i Rozwoju (LIDER/35/0127/L9/17/NCBR/2018), and Fundacja na rzecz Nauki Polskiej (TEAMTECHPOIR.04.04.00-00-210C/16-00).